\documentclass[doublecol]{epl2} 

\usepackage{bm}
\usepackage{chemarr}
\usepackage{amsmath}
\usepackage{color}
\usepackage{amssymb}
\usepackage[caption=false]{subfig}

\newcommand{\bea}{\begin{eqnarray}}
\newcommand{\eea}{\end{eqnarray}}

\title{Dynamically Emergent Correlations}
\shorttitle{Dynamically Emergent Correlations} 
\author{Satya N. Majumdar\inst{1} \and Gr\'egory Schehr\inst{2}}
\shortauthor{}

\institute{                    
  \inst{1} LPTMS, CNRS, Univ.  Paris-Sud,  Universit\'e Paris-Saclay,  91405 Orsay,  France \\
  \inst{2} Sorbonne Universit\'e, Laboratoire de Physique Th\'eorique et Hautes Energies, CNRS UMR 7589, 4 Place Jussieu, 75252 Paris Cedex 05, France
}

\abstract{In this {\it perspective article}, we discuss the scenario of dynamically emergent 
correlation (DEC) arising in classical and quantum noninteracting systems when they are 
subjected to a common fluctuating stochastic environment. The key property of such systems 
is that the strong correlations between different particles emerge from the dynamics and 
not from built-in interactions. In many cases, these strong correlations persist even at 
long times in the stationary state. Computing observables explicitly for such strongly 
correlated states in general is very hard. Remarkably, the stationary states in several 
models of DEC exhibit an interesting analytical structure that allows to compute physical 
observables, despite being strongly correlated. Recent experiments on trapped colloidal 
particles have established that these DEC in the stationary state can in fact be measured. 
DEC is a rapidly emerging domain of strongly correlated out-of-equilibrium statistical 
physics, with both theoretical and experimental, as well as classical and quantum, components.
}

\begin{document}

\maketitle

Dynamically emergent correlations (DEC) have recently appeared as a very interesting 
phenomenon whereby different degrees of freedom in a system, which are otherwise 
noninteracting, get correlated with increasing time due to the fact that they share a common 
environment that fluctuates stochastically and independently of the particles 
\cite{BLMS23,BLMS24,BKMS24,SM2024,MBMS25,MMS25,KMS25,MKMS25,BMS25,BM25,BMS26,VR25,BCKMPS25,MMS26}. 
We call these correlations {\it emergent} because they are absent if this environment does 
not fluctuate with time or even when it evolves deterministically. Thus the ``stochastic'' 
component of the dynamics of the environment is essential for the emergence of these 
correlations.

\begin{figure}[t]
\includegraphics[width = \linewidth]{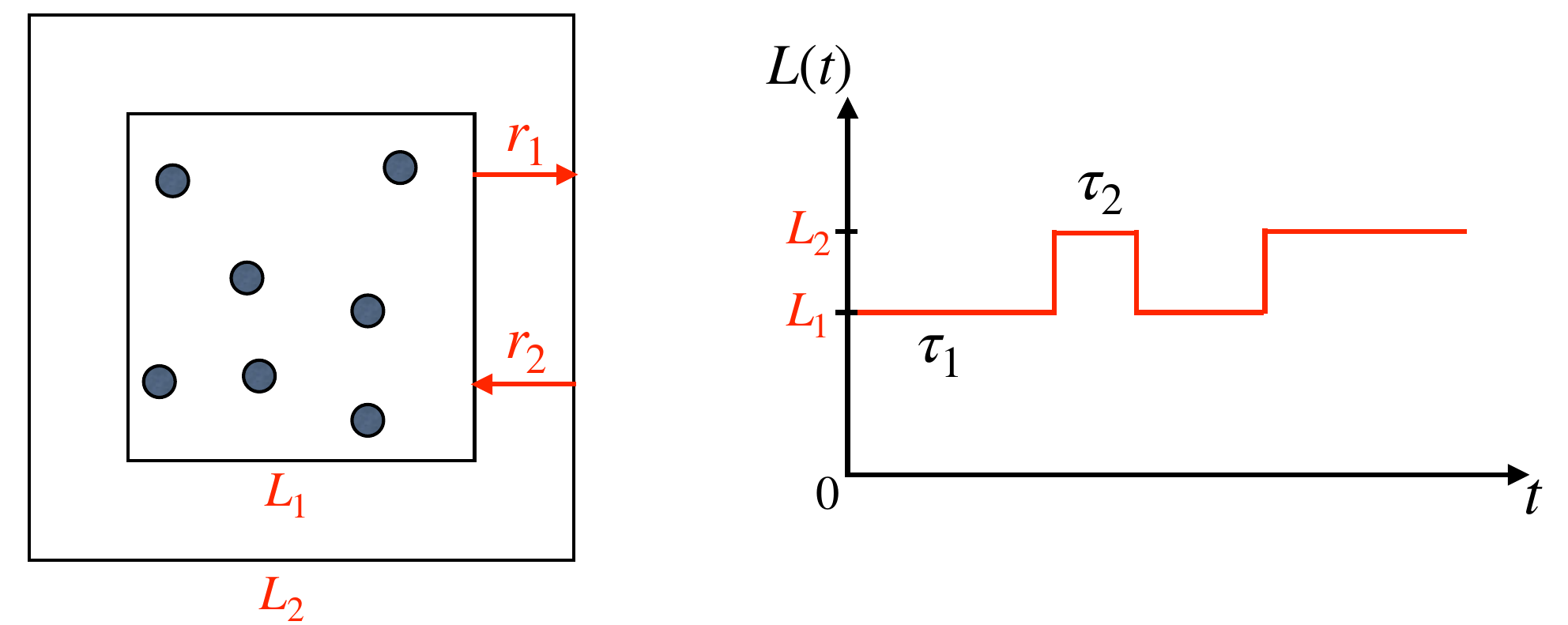}
\caption{The left panel shows a schematic picture of an ideal gas of Brownian particles 
in a two-dimensional box of linear size $L(t)$ which switches between $L_1$ and $L_2$ 
in a noisy dichotomous (telegraphic) fashion, as shown in the right panel. 
The intervals between successive switches are drawn 
alternatively from exponential distributions $p_1(\tau) = r_1\, e^{-r_1 \tau}$ 
and $p_2(\tau) = r_2\,e^{-r_2 \tau}$ respectively. Here, $r_1$ and $r_2$ represent the 
Poissonian rates with which the box size switches from $L_1$ to $L_2$ and the reverse.}\label{Fig_Intro}
\end{figure}
Such emerging correlations between different independent constituents were first observed 
experimentally in a very different context by Huygens, who besides being a great mathematician, was also an expert 
clock maker in Europe in the $17^{\rm th}$ century \cite{Huygens}. He performed a  
simple experiment where he hung two independent pendulums from a wooden beam bridging two 
chairs. One pendulum was kicked gently. While it starts swinging, the other pendulum 
remained static for a while, but afterwards the two pendulums started swinging 
synchronously, indicating the presence of correlations. 
However, the physical mechanism of the emergent correlations in Huygens' s setting is quite 
different from the DEC that we discuss here. In Huygens's case, when one pendulum (particle) 
is perturbed externally, the other one moves due to the nonlocal response mediated via the beam
connecting the pendulums (particles). This is fundamentally different from our case in two ways: 
(i) we do not have any information on the positions of the individual particles and the drive is imparted 
on the system as a whole, which all particles share, (ii) there is no physically connecting link between any pair of particles, unlike
the rigid wooden beam in Huygens's setup. Furthermore, in the Huygens's setup, one particle is kicked only once, while in our case
the perturbation due to the fluctuations of the environment is continuous in time. However, 
quite importantly, in our setup the particles have no effect 
on the dynamics of the environment -- instead the environment has its own stochastic motion driven from outside 
and, as we will see, this is sufficient to induce correlations between the particles.

As a simple setting for DEC, consider an ideal gas of $N$ noninteracting 
Brownian particles confined in a $d$-dimensional box of linear size $L(t)$.  The size $L(t)$ 
of the box can be varied with time $t$ in a controlled way. For example, in $1d$, one can 
put two pistons on opposite sides of the box and either push or pull the piston to change 
$L(t)$. We consider a very simple protocol. We start the system with box size $L_1 > 0$ and 
wait for an exponentially distributed random time $\tau_1$, drawn from $p_1(\tau) = r_1\, 
e^{- r_1 \tau}$, during which the particles undergo independent diffusion with reflecting 
boundary conditions at the box walls. After this exponential random time, we instantaneously 
change the box size to $L_2 > L_1$. We then again wait for another exponentially distributed 
random time $\tau_2$ drawn this time from $p_2(\tau) = r_2\, e^{- r_2 \tau}$ during which 
the particles again diffuse independently in the box of size $L_2$ with reflecting boundary 
conditions at the walls. Following this, we instantaneously push the box size back to $L_1$ 
and repeat the process. The box, here, represents the environment, whose linear size $L(t)$ 
undergoes a noisy dichotomous telegraphic evolution in time (see Fig. \ref{Fig_Intro}). The 
particles, with no direct interactions between them, are simply driven by this fluctuating 
box. As a little detail, one may wonder what happens, when the box size is reduced from 
$L_2$ to $L_1$, to the particles that lie between $L_1$ and $L_2$. One can model this in 
different ways. For example, one simple choice would be to make those particles stick to the 
pushed wall. Other choices are also possible, but these details do not make any difference 
in the qualitative understanding of the underlying physics.

Given this simple setting with an ideal gas driven by the fluctuating box size, our main 
goal is to investigate whether such a fluctuating environment induces correlations between 
the noninteracting particles as time increases. In particular, does the system reach a 
stationary state at long times and do these emergent correlations persist all the way up to 
the stationary state? More precisely, one would like to compute the joint probability 
distribution function (JPDF) $P({\bf x}_1, {\bf x}_2, \cdots, {\bf x}_N; t)$ of the 
positions of the particles ${\bf x}_1, {\bf x}_2, \cdots, {\bf x}_N$ at time $t$. If this 
JPDF factorizes into individual marginal distributions, then the gas remains ideal, 
indicating the absence of correlations between the particles. However, it turns out not to 
be the case, as we will see. Moreover, we will see that the JPDF becomes 
time-independent as $t \to \infty$, indicating the presence of a nonequilibrium stationary 
state (NESS) with strong correlations emerging between every pair of particles. The 
stationary state is nonequilibrium since the dynamics of the box breaks the detailed balance 
condition, leading to nonzero probability currents in the configuration space.

Despite the simplicity of this model, it turns out to be rather hard 
to solve analytically. Instead, it is easier to investigate a particular limit 
that still captures the main physical phenomenon, while remaining solvable 
analytically. The idea is to consider the limit $L_2 \to \infty$, $L_1 \to 0$, $r_1 \to 
\infty$ and $r_2 = r$ fixed. This limiting situation has only one free parameter $r$. In 
this limit, the dynamics of the particles reduce to the following. All the particles start 
at the origin (since $L_1 = 0$) and instantaneously start diffusing (since $r_1\to \infty$) 
independently in free space (since $L_2 \to 
\infty$). After free diffusion during a random time drawn from $p(\tau) = r\,e^{-r \tau}$ 
(since $r_2 = r$), the box size is squeezed to the origin instantaneously and released 
immediately (since $r_ 1 \to \infty$). This corresponds precisely to $N$ Brownian 
motions that diffuse independently starting from the origin and are reset simultaneously 
to the origin with rate $r$. This is a multi-particle generalisation of single 
particle resetting that has been studied over the last decade quite extensively, with many 
applications~\cite{EM11a,EMS20}. This multi-particle generalisation with simultaneous 
resetting was studied recently in Ref. \cite{BLMS23}, where the JPDF $P({\bf x}_1, {\bf 
x}_2, \cdots, {\bf x}_N; t)$ was computed explicitly at all times and the existence of DEC 
was first established. Here, we briefly recall this result and, for simplicity, 
focus on $d=1$.
\begin{figure}[t]
\includegraphics[width = \linewidth]{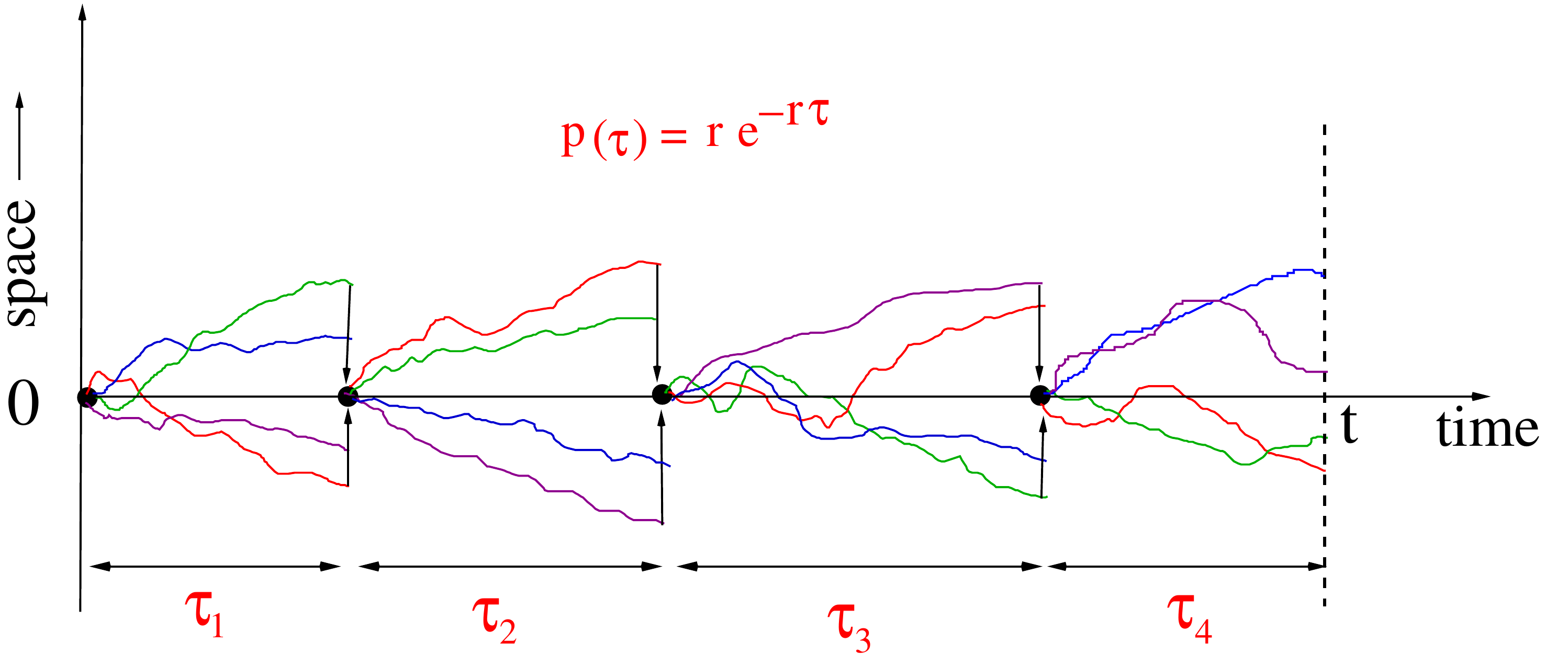}
\caption{Schematic trajectories of $N$ independent Brownian motions in one-dimension that start and reset simultaneously at the origin with rate $r$. This corresponds to the limiting case $L_1 \to 0$, $L_2 \to \infty$, $r_1 \to \infty$ and $r_2=r$ of Fig. \ref{Fig_Intro} in the $1d$ case. Here the $\tau_i$'s denote the intervals between successive resettings and are distributed independently, each drawn from $p(\tau) = r\,e^{-r \tau}.$}\label{Fig_resetting}
\end{figure}

Using the shorthand notation $\vec{x} \equiv \{x_1, x_2, \cdots, x_N\}$, the JPDF 
$P_r(\vec{x};t)$ of the positions of the particles can be computed using a renewal property 
of the dynamics~\cite{BLMS23}
\begin{equation} \label{eq:renewal}
P_r(\vec{x};t) = e^{-r\, t}  \prod_{i=1}^N G_0(x_i,t) + r\int_0^t d\tau \, e^{-r \tau} \,\prod_{i=1}^N G_0(x_i, \tau) \;, 
\end{equation}
where $G_0(x,\tau) = e^{-\frac{x^2}{4 D\,\tau}}\,/\sqrt{4 \pi D\, \tau}$ is the bare 
(reset-free) single particle propagator. This result can be understood very simply. In a 
given trajectory of the process up to time $t$, there can be either $0$ or at least one 
resetting event. In the former case, the particles evolve freely and independently, each 
with the propagator $G_0(x,t)$. In this case, the probability of no resetting in the time 
interval $[0,t]$ is simply $e^{-r\,t}$ since the resetting events are Poissonian with rate 
$r$. This then gives the first term on the right hand side (RHS) of Eq. (\ref{eq:renewal}). 
When there are one or more resettings in the trajectory, one needs to first identify
the epoch $t-\tau$ at which the last resetting occurred before $t$. Then the system 
undergoes free diffusion of duration $\tau$ in the interval $[t-\tau,t]$, as given by the 
product $\prod_{i=1}^N G_0(x_i, \tau)$ in Eq. (\ref{eq:renewal}). The probability that the 
last resetting occurs at $t-\tau$ is simply $r \,d\tau$. This has to be followed by no 
resetting during $\tau$, whose probability is simply $e^{-r\tau}$. Taking the product of 
these three independent events and integrating over $\tau \in [0,t]$ gives the second term 
in the RHS of Eq. (\ref{eq:renewal}). One can easily check that $\int 
d\vec{x}\,P_r(\vec{x};t) = 1$, ensuring the normalisation at all times~$t$. We note that 
$P_r(\vec{x};t)$ can also be interpreted as the position distribution of a single Brownian 
particle undergoing stochastic resetting to the origin at rate $r$, but in the 
$N$-dimensional space, which was studied in Ref. \cite{EM14} in terms of the radial 
coordinate $R = \sqrt{x_1^2+x_2^2 + \cdots + x_N^2}$. However, here, our goal is to study 
the correlations between different particle coordinates $x_i$ and $x_j$ with $i \neq j$, and 
also other associated observables, such as the gaps between two consecutive particles on the 
line. In terms of the components of a $N$-dimensional vector, these observables do not have 
any natural physical meaning, hence we prefer to interpret $x_i$'s as the positions of $N$ 
particles on the line, rather than the components of a single vector in the $N$-dimensional 
space.

By looking at Eq. (\ref{eq:renewal}), it is obvious that 
$P_r(\vec{x};t)$ does not factorize, indicating that the positions of the particles 
become correlated at any time $t > 0$. 
To measure these correlations, it would be natural to first consider the standard connected correlation function $\langle x_i(t) x_j(t)\rangle - \langle x_i(t) \rangle\langle x_j(t)\rangle$.
However, since the JPDF $P_r(\vec{x};t)$ in \eqref{eq:renewal} has the $x_i \to -x_i$ symmetry, it follows that this standard correlation function vanishes identically at all times $t$. Hence, to detect the nonzero correlations, we need to probe higher order correlation functions. The simplest nonzero higher order connected correlation function is the covariance between $x_i^2(t)$ and $x_j^2(t)$ for $i \neq j$~\cite{BLMS23,BLMS24} 
\begin{equation} \label{eq:def_C2}
{\cal C}_2(t) = {\rm Cov}(x_i^2(t),x_j^2(t)) = \langle x_i^2(t) x^2_j(t)\rangle - \langle x^2_i(t) \rangle\langle x^2_j(t)\rangle \;.
\end{equation}
Given the JPDF in Eq. (\ref{eq:renewal}), it is clear that ${\cal C}_2(t)$ does not depend on the particle indices $i$ and $j$, as long as $i \neq j$. 
Sometimes it is also useful to study the dimensionless covariance function ${\rm Cov}(x_i^2(t),x_j^2(t))/{\rm Var}(x_i^2(t))$ which then reads
\bea \label{eq:def_At}
A(t) = \frac{\langle x_i^2(t) x^2_j(t)\rangle - \langle x^2_i(t) \rangle\langle x^2_j(t)\rangle}{\langle x_i^4(t)\rangle - \langle x^2_i(t) \rangle^2} \;.
\eea  
It is easy to show that $A(t)$ lies in the interval $[0,1]$. When $A(t) \to 0$, the system is completely uncorrelated, while the limit $A(t) \to 1$ corresponds to the maximally correlated case when $x_i(t)=x_j(t)$~\cite{BM25}. While $A(t)$ was computed exactly in the stationary limit $t \to \infty$ in Ref. \cite{BLMS23}, it can also be computed exactly at all times $t$ -- see Refs. \cite{MBMS25,BM25}. In fact, $A(t)$ can be expressed as a function of the dimensionless time $z=r\,t$ as  
\bea \label{eq_aofz}
A(t) = a(z=r\,t) = \frac{1-2 z \,e^{-z} - e^{-2z}}{5-(4+6z)\,e^{-z} - e^{-2z}} \;.
\eea

It starts as $a(z) \approx z/6$ as $z \to 0$, grows monotonically with increasing $z = r\,t$ 
and finally saturates to a nonzero value $a(z \to \infty) = 1/5$. This observable $A(t)$ 
quantifies the DEC and clearly demonstrates that the correlations between the particles 
emerge dynamically, grow with time and persist all the way to the stationary state, where 
the correlations become the strongest. Since these correlations do not depend on the indices 
$(i,j)$ of the particles, it represents an all-to-all correlation (as in mean-field models 
of statistical physics). Moreover this represents an attractive correlation, which is 
natural since simultaneous resetting tends to bring the particles together to the origin. We 
also note that, since there is no finite correlation length, this system represents a {\it 
strongly correlated} system. For such strongly correlated systems, computing the statistics 
of physical observables (to be defined more precisely later) are usually hard, even if one 
knows the JPDF of the positions explicitly. A classical example of that is the statistics of 
the eigenvalues of a random Gaussian matrix where every pair of eigenvalues repel each other 
\cite{Mehta,For}. To compute observables like the density of eigenvalues, the distribution 
of the largest eigenvalue or the spacing distribution between the eigenvalues took many 
years~\cite{Mehta,For}. In our system which is also strongly correlated, these 
observables are actually much easier to compute, due to a a particular mathematical 
structure of the JPDF in the stationary state~\cite{BLMS23}, as we will see below.

\begin{figure}[t]
\centering
\includegraphics[width = 0.8\linewidth]{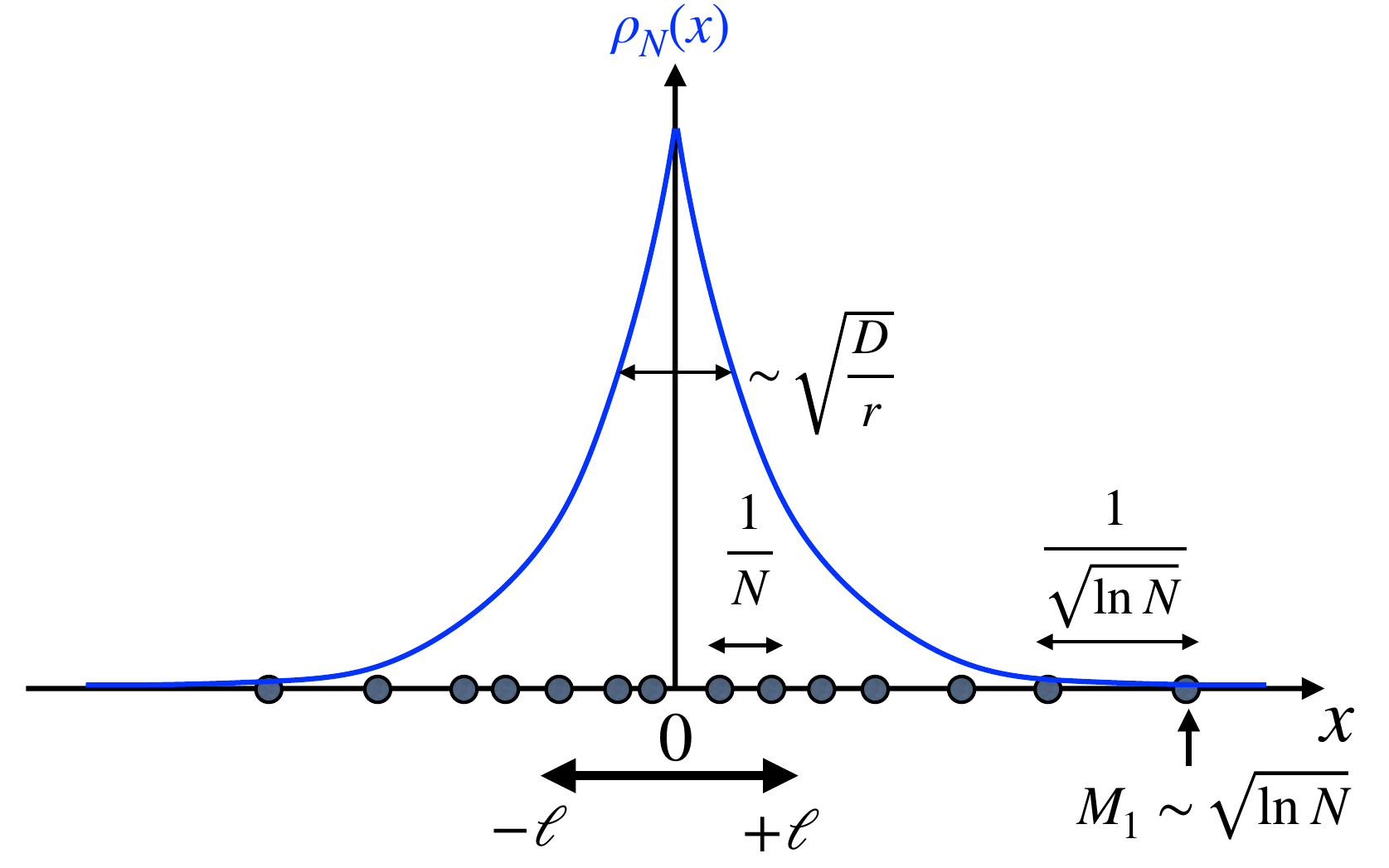}
\caption{The blue line shows the average density profile in Eq.~(\ref{eq:density}).}\label{Fig_density}
\end{figure}

Having established that the correlations grow with time due to the fluctuating environment, we now focus on the stationary state in the long time limit $t \to \infty$, where the JPDF in Eq. (\ref{eq:renewal}) becomes~\cite{BLMS23}
\bea \label{eq:stationary}
P^{\rm st}_r(\vec{x}) =  r\int_0^\infty d\tau \, e^{-r \tau} \,\prod_{i=1}^N G_0(x_i, \tau) \;.
\eea
Given this stationary JPDF, one would like to compute several observables. 
This represents a one-dimensional gas of $N$ strongly correlated particles and our next goal 
is to compute, form this stationary measure, several macroscopic and microscopic observables associated to this gas. Since the gas is not homogeneous in space, the first natural macroscopic observable to probe is the average density profile of the gas defined as $\rho_N(x) = (1/N) \sum_{i=1}^N \langle \delta(x-x_i)\rangle$ which counts the average fraction of particles per unit length at $x$, with the normalization $\int \rho_N(x) \, dx =1$. From the JPDF in \eqref{eq:stationary}, it is easy to compute the average density. It turns out to be independent of $N$ and is given by~\cite{BLMS23}
\bea \label{eq:density}
\rho_N(x) = \frac{1}{2} \sqrt{\frac{r}{D}}\, e^{-|x| \sqrt{\frac{r}{D}}} \;.
\eea
The density profile is thus non-Gaussian with a cusp at $x=0$ (see Fig. \ref{Fig_density}). 
The peak at $x=0$ indicates that the gas is more dense near the origin (this is generated by 
the simultaneous resetting of the gas to the origin with rate $r$). The density decays 
exponentially over a length scale $\xi = \sqrt{D/r}$ that characterizes the typical distance 
travelled by a single particle between two consecutive resettings. The gas becomes sparser 
and sparser as one goes away from the origin. Having an idea of the macroscopic density 
profile, we now probe the gas at a more microscopic level. For this it is convenient to 
first order the positions of the particles $\{x_i\} \to \{M_i\}$ with $M_1 > M_2> \cdots> 
M_N$. For example, $M_1$ corresponds to the position of the rightmost particle in the gas -- 
see Fig. \ref{Fig_density} -- while $M_k$ represents the position of the $k$-th rightmost 
particle from the right. Another microscopic observable is the 
spacing $d_k = M_k - M_{k+1}$ between the $k$-th and the $(k+1)$-th particle. A third 
natural observable is the full counting statistics (FCS), denoting the distribution of the 
number of particles $N_{\ell}$ in the interval $[-\ell,+\ell]$ around the origin, as
indicated in ig. \ref{Fig_density}.

We now discuss how these observables can be computed in the stationary state by exploiting a 
particular mathematical structure of the stationary JPDF in Eq. (\ref{eq:stationary}). From 
this equation, we see that, if we fix $\tau$, then we have a set of independent Gaussian 
variables with the JPDF $\prod_{i=1}^N G_0(x_i, \tau)$. Exploiting this IID
structure (i.e., an ideal gas) at fixed $\tau$, one can easily compute the 
observables mentioned above. 
Once we have computed the statistics of these observables 
for fixed~$\tau$, we just need to average over the distribution of the random variable 
$\tau$, which in this case is a simple exponential $p(\tau) = r\,e^{-r\tau}$. Thus the JPDF 
has the following general mathematical form
\bea \label{P_CIID}
P_{\rm CIID}(\vec{x}) = \int du\, h(u)\, \prod_{i=1}^N p(x_i \vert u) \;,
\eea   
where $u$ is the conditioning random variable, distributed via the PDF $h(u)$, while $p(x_i \vert u)$ denotes the conditional PDF of $x_i$, given $u$. We call this structure (\ref{P_CIID}) conditionally independent and identically distributed (CIID) \cite{BLMS23,BLMS24}. In the context of the resetting gas, the conditioning variable $u$ is simply the time $\tau$ representing the time since the last resetting and $h(u) \to p(\tau) = r\,e^{-r\tau}$. We will see later that a similar CIID structure emerges in several problems not necessarily related to resetting. In fact, in many situations there can be a hidden CIID structure and the challenge, then, is to identify the conditioning variable $u$ and its distribution $h(u)$.

Exploiting this CIID structure for the resetting gas, the microscopic observables mentioned 
above were computed analytically in Refs. \cite{BLMS23,BLMS24}. We do not provide the 
details here, but just summarise the salient features. Since the gas becomes sparser away 
from the origin, $M_1 \gg 1$ in the large $N$ limit. In fact, it was shown that $M_1 \sim 
\sqrt{\ln N}$ for large $N$. The spacing $d_k = M_k - M_{k+1}$ typically scales as $1/N$ 
when $k = O(N)$ (we call this region the ``bulk'', which is around the origin), while, near 
the ``edge'' where $k=O(1)$, the typical spacing scales as $1/\sqrt{\ln N}$ -- see Fig. 
\ref{Fig_density}. Surprisingly the scaled distribution of the $k$-th maximum, in the large 
$N$ limit, becomes universal, i.e., independent of the index $k$, and is given by a simple 
scaling function~\cite{BLMS23,BLMS24}. Similarly, the scaled gap has a nontrivial 
distribution (with a stretched exponential tail), which is also universal, i.e., independent 
of $k$. The FCS, also, turns out to be highly nontrivial with a nonmonotonic 
shape~\cite{BLMS23,BLMS24}. The FCS can be regarded as a linear statistics of the form
${\cal L}_N = \sum_{i=1}^N f(x_i)$ where $f(x) = {\cal I}_{[-\ell, +\ell]}(x)$ 
is an indicator function. For general $f(x)$, an information theoretic bound on the relative 
fluctuations of ${\cal L}_N$ was derived recently for stationary states with a 
CIID structure \cite{Olsen26}.

The results above were derived for a resetting noninteracting 
Brownian gas. However, the renewal structure in Eq. (\ref{eq:renewal}), and consequently the 
CIID structure of the NESS in (\ref{eq:stationary}), are very general and hold for any 
resetting noninteracting gas with arbitrary stochastic dynamics of each individual particle, 
not necessarily Brownian. The only thing that comes as an input in the CIID structure is the 
free propagator $G_0(x,\tau)$ of the gas. In fact, all these observables have been computed 
for various dynamics of the resetting gas, such as ballistic particles with random initial 
velocity and L\'evy flights~\cite{BLMS24}. In principle, this method can be used for other 
types of dynamics of particles between resettings, including for instance the run and tumble 
dynamics. This remains an interesting open problem for future studies. 

Another interesting generalisation considers simultaneous resettings but the time interval 
between two successive resettings is not a pure exponential form, but with arbitrary 
$p(\tau)$. This is called non-Poissonian resetting because one can not think of it as a 
continuous process with some rate $r$. Nevertheless the renewal structure in Eq. 
(\ref{eq:renewal}) can be easily generalised to the non-Poissonian case and generically, one 
again finds a NESS with a CIID structure~\cite{MBMS25}. Let us remark here that if 
the resetting events occur stroboscopically with period $T$, i.e., if $p(\tau)= 
\delta(\tau-T)$, then the system will reach a time periodic stationary state (TPSS) rather than 
a true NESS. In this TPSS, at any given instant, it is easy to see that the JPDF
factorizes completely, indicating the absence of correlation between 
particles. Thus `stochasticity' in the fluctuating environment is necessary for DEC. Finally, this model of simultaneous resetting of $N$ particles was 
recently generalized to resettings with memory~\cite{BM25}. In this last model, 
there is neither a NESS nor a TPSS. Nevertheless, the time-dependent JPDF still has a CIID structure at any time $t$~\cite{BM25}.

\noindent{\it Harmonically trapped gas with switching stiffnesses.} So far we have discussed the limiting case of the ideal Brownian gas in the box problem (see Fig. \ref{Fig_Intro}) when $L_1\to 0$, $L_2 \to \infty$, $r_1 \to \infty$ and $r_2 = r$. It turns out that solving this problem for general $L_1, L_2$, $r_1$ and $r_2$ is very hard. However, a close cousin of this model can be solved exactly~\cite{BKMS24}. In this version one considers $N$ noninteracting Brownian particles diffusing in a harmonic potential $V(x) = \frac{1}{2} \mu(t)\, x^2$ whose stiffness $\mu(t)$ changes dichotomously between two values $\mu_1$ and $\mu_2$ (with $\mu_1 > \mu_2$) with rate $r_1$ (from $\mu_1$ to $\mu_2$) and $r_2$ (from $\mu_2$ to $\mu_1$). Here the typical length scale over which the particles are confined is $1/\sqrt{\mu(t)}$, which plays the analogous role of the length $L(t)$ in the box problem. Moreover, as we will see shortly, this model is easier to access experimentally via laser
trapping of colloidal particles~\cite{BCKMPS25}. The first hurdle in solving this problem is that it does not have the simple renewal structure as in Eq. (\ref{eq:renewal}), valid for the limiting box problem. Hence, it is not obvious whether there is a CIID structure for the JPDF in the stationary state. Instead, one can obtain the JPDF by directly solving the associated Fokker-Planck equations~\cite{BKMS24}. Skipping details, we find that the JPDF has again a CIID structure as in Eq. (\ref{P_CIID}) with
\bea \label{eq:px_switch}
p(x_i \vert u) = \frac{e^{-\frac{x_i^2}{2 V(u)}}}{\sqrt{2 \pi V(u)}}  \quad , \quad V(u) = D \left(\frac{u}{\mu_2} + \frac{1-u}{\mu_1} \right) \;,
\eea
and $h(u) = A \,u^{R_1-1}(1-u)^{R_2-1} \left[\frac{1-u}{\mu_1} + \frac{u}{\mu_2} \right] \;, u\in[0,1]$. Here $R_1 = r_1/(2\mu_1)$, $R_2 = r_2/(2\mu_2)$ and the amplitude $A$ normalises $\int_0^1 h(u) du = 1$. The random variable $u \in [0,1]$ can be interpreted as the effective fraction of time a particle spends in phase $2$ and $V(u)$ is the average variance of a single Ornstein-Uhlenbeck (OU) particle in the switching trap. Let us make two observations: (i) identifying the conditioning variable $u$ in the CIID structure in Eq. (\ref{P_CIID}) was not apparent to start with and (ii) computing the PDF $h(u)$ of $u$ is also nontrivial. For an alternative and somewhat more transparent derivation of this result, we refer the reader to Ref. \cite{BKMS24}. Once this CIID structure is revealed, computing the observables is rather straightforward~\cite{BKMS24}, as 
in the previous model.

\noindent{\it Experimental realisation.} This system of noninteracting Brownian particles in a switching harmonic trap has recently been realised in experiments on trapped colloidal particles in a liquid (water) \cite{BCKMPS25}. In this setup the particles were not localized in one single trap, as in the model described above, but rather there were $N$ separate harmonic traps, each containing a single particle with two independent $x$ and $y$ coordinates. The stiffnesses of different traps were synchronously modulated between two values $\mu_1$ and $\mu_2$. In the experiment, a maximum of four traps could be used. Unlike in the model where the particles are noninteracting, in the experimental system the particles are embedded in the water and there is always a long-range hydrodynamic interaction between the particles. The point of performing this experiment was not just to verify the theoretical predictions (which was already done by numerical simulations), but rather to investigate to what extent the theoretical predictions based on noninteracting particles continue (or not) to hold in the presence of long range hydrodynamic interactions.  
To be more precise, the NESS correlation function ${\cal C}_2(t \to \infty)$ in Eq. \eqref{eq:def_C2} can be nonzero due to two different sources: (i) the DEC where the correlations are generated dynamically due to the simultaneous switching of the trap and (ii) the long-ranged hydrodynamic interactions mediated by the fluid in which particles are embedded. To our surprise, we found that, even though hydrodynamic interactions are unambiguously present (as verified by independent measurements of other observables \cite{BCKMPS25}), the measured stationary correlation function in the NESS, namely ${\cal C}_2(t \to \infty)$ (appropriately normalised) matched extremely well with the noninteracting theory (for details see Ref. \cite{BCKMPS25}). This led to the conclusion that, while the hydrodynamic interactions may affect some observables, the correlation function ${\cal C}_2(t \to \infty)$ is completely dominated by the DEC, which overwhelms the hydrodynamic interactions. A similar conclusion was also reached in   
a different experiment \cite{VR25}, where the particles were simultaneously reset mechanically using optical tweezers.  

\noindent{\it Tuning the strength of correlations in the NESS.}  In both models discussed above, we have seen that the simultaneous resetting (in the limiting box problem) and the synchronous switching of the harmonic potential, a NESS is generated in the long time limit with all-to-all attractive correlations between the particles. A natural question is to ask if one can tune the strength of these correlations in the NESS by relaxing the hard constraint of ``simultaneous'' nature of the resetting or the fully synchronised nature of the harmonic trap. One simple way to relax these hard constrains is to study the so called ``batch resetting'' introduced recently in Ref. \cite{MMS26}. Consider, for simplicity, the first problem of $N$ noninteracting Brownian motions being simultaneously reset to the origin with rate $r$. Instead of resetting all $N$ particles simultaneously, one can study what happens if only $m$ particles, chosen randomly out of $N$, are reset together with rate $r$, while the rest of the $N-m$ particles continue their Brownian trajectories. The parameter $m$ allows to interpolate between the uncorrelated case ($m=1$) and the strongly correlated simultaneous resetting case ($m=N$). In this case, for $1<m<N$, the stationary JPDF does not have any apparent CIID structure. Nevertheless, the two-point correlation function ${\cal C}_2(t \to \infty)$ can be computed exactly revealing strong tuneable correlations \cite{MMS26}. However, computing other observables such as the extreme value statistics, the spacing distribution between particles or the full counting statistics for any $1<m<N$ remains a challenging open problem. It will be interesting to see whether these tunable correlations can be measured in experiments on trapped colloids mentioned above \cite{BCKMPS25}. 

\noindent{\it DEC for classical protocols going beyond dichotomous driving.} So far, the 
stiffness of the harmonic well was changing dichotomously between two values $\mu_1$ and 
$\mu_2$ with exponentially distributed time intervals between two successive switches.  It 
is natural to ask: why dichotomous noise (with two values only) 
and why not some continuous stochastic $\mu(t)$? The 
latter turns out to be very hard to solve and it remains an open problem. However, there is 
another version of the model that can be solved for a continuous driving force. Instead of 
changing the stiffness $\mu(t)$ of the trap, we drive the center of the trap by an external 
stochastic force. More precisely, we consider $N$ independent OU processes in a harmonic 
potential with stiffness $\mu$ and the trap center located at $z(t)$ which fluctuates
stochastically. Thus the positions $x_i(t)$ of the particles evolve via the 
overdamped Langevin equation \cite{SM2024}
\bea \label{eq:OU_z}
\frac{dx_i}{dt} = - \mu (x_i - z(t)) + \sqrt{2D}\,\eta_i(t) \;, 
\eea 
where $\eta_i(t)$'s are independent Gaussian white noises of zero mean and the correlator 
$\langle \eta_i(t) \eta_j(t') \rangle = \delta_{ij}\,\delta(t-t')$. Here $z(t)$ represents 
the stochastic drive, i.e., the fluctuating environment . For example, $z(t)$ could be just 
a white noise or a dichotomous noise. The stationary JPDF in this model can be computed 
exactly and it shows a CIID structure~\cite{SM2024}
\bea \label{eq:st_z}
P^{\rm st}(\vec x) = \int du \, h(u) \prod_{i=1}^N \sqrt{\frac{\mu}{4 \pi D}} e^{- \mu\frac{(x_i-u)^2}{4D}} \;,
\eea
where the conditioning variable $u$ evolves via ${du}/{dt} = - \mu u + \mu\,z(t)$ and $h(u)$ 
represents its stationary distribution. This CIID structure can then 
be exploited to compute the standard observables mentioned before, leading to another interesting 
strongly correlated and yet solvable system~\cite{SM2024}. It would be interesting to find 
other solvable models of independent particles with a shared driving force that is 
also continuous in time.

Another interesting angle is to consider a fluctuating driving force with both spatial and 
temporal degrees of freedom. For example, consider $N$ independent overdamped particles in 
one dimension in the presence of a potential $h(x,t)$ evolving via the noisy gradient 
descent dynamics \cite{NBM05,NMB06}, $ {d x_i}/{dt} = - \Gamma \partial_{x_i} h(x_i,t) + 
\sqrt{2D}\, \eta_i(t)$ where $\eta_i(t)$'s are independent Gaussian white noises as before 
and $h(x,t)$ represents a fluctuating height of a potential landscape in which the particles 
are moving. A natural choice for the dynamics of this height field can for instance be the 
Kardar-Parisi-Zhang (KPZ) equation $\partial_t h(x,t) = \nu \partial^2_x h + \lambda 
(\partial_x h)^2 + \zeta(x,t)$, where $\zeta(x,t)$ is a white noise with zero mean and a 
correlator $\langle \zeta(x,t) \zeta(x',t') \rangle = 2 \delta(x-x')\delta(t-t')$. This 
model was studied in Refs. \cite{NBM05,NMB06} both numerically and analytically in some 
limiting cases. It was found that, in the stationary state, the particles show strong 
clustering in space, indicating that they do get strongly correlated, being driven by the 
same fluctuating field. When the height field fluctuates slowly compared to the time scale 
of the particle's motions, the correlation function can be computed analytically via a 
mapping to the well known Sinai model of disordered systems~\cite{NMB06}.  However, when the 
two timescales are of the same order, there is no known analytical result for the 
correlation function between particles in the stationary state, even in one-dimension, and 
this remains a challenging open problem.

\noindent{\it DEC for driven quantum particles.} So far we treated mostly {\it classical} 
independent particles subject to a common external stochastic drive and in most cases we have 
found that the stationary state exhibits a CIID structure, which is sometimes explicit and 
sometimes hidden. It is natural to ask if there is an equivalent DEC for particles that 
undergo unitary quantum evolution, instead of classical dynamics: does the DEC exist for the 
quantum evolution? This brings us naturally to the idea of quantum resetting, which is 
becoming rather popular 
recently~\cite{MSM18,RTLG18,PCML21,MCPL22,Nav18,KM23,YB23,Mor24,Barkai25,Jaf26,Mor26}. The 
idea behind quantum resetting is very simple: it consists of a dynamics which has a 
deterministic quantum component and a classical stochastic component~\cite{MSM18,RTLG18}. 
Consider a quantum system in a state $\Psi(t)$ that evolves by the following dynamics. In a 
small time $dt$, the state evolves to $\Psi(t+dt)$~\cite{MSM18}
\bea \label{eq:quantum} 
\Psi(t+dt) = 
\begin{cases}
&\hspace*{-0.2cm}\left(1 - \frac{i}{\hbar} \hat H \, dt\right)\Psi(t)  \;, \; {\rm with \; prob.}\; 1 - r \,dt \\
&\hspace*{-0.2cm}\Psi(0) \;, \; \hspace*{2cm}{\rm with \; prob.} \; r \,dt \;. 
\end{cases} 
\eea
Here $r$ represents the resetting rate. Whether DEC occurs between independent quantum degrees of freedom when they are subject to simultaneous quantum resetting is then an interesting question and is the analogue of the classical DEC discussed before. In fact, this question of the emergence of correlations between independent quantum degrees of freedom was first addressed in Ref. \cite{MCPL22} where the authors studied two noninteracting Ising spins in a transverse field and subject them to simultaneous quantum resetting after a unitary evolution up to a random time. It was shown that the system reaches a stationary state at long times, with nonzero correlations~\cite{MCPL22}. 

Going beyond two degrees of freedom, a solvable example with simultaneous quantum resetting consists of 
$N$ noninteracting bosons in a one-dimensional harmonic potential. The system is prepared in the ground state with the harmonic potential centred at $+a$ and then evolved unitarily with a potential centred at $-a$. The evolution continues up to a random time, drawn from a distribution $p(\tau) = r\,e^{-r \tau}$. After this random time, the system is reset back to the ground state of the system with the potential centered at $+a$. 
This plays the analogous role of simultaneous resetting in classical systems. 
Then the whole cycle is repeated again and again. We ask whether the quantum probability density encoded by $|\Psi(\vec x,t)|^2$, where $\Psi(\vec x,t)$ is the many-body wave function at time $t$, evolves to a stationary state, i.e., becomes independent of time as $t \to \infty$. Indeed, it was shown in Ref. \cite{KMS25} 
that the stationary quantum probability density also has a CIID structure. Consequently, all the observables mentioned in the classical case could also be computed exactly in this quantum case \cite{KMS25}. One could consider alternative protocols for the quantum evolution where, instead of the trap center, we switch the frequency of the trap~\cite{MKMS25}. In this case, one finds again that the stationary quantum JPDF $|\Psi(\vec x,t \to \infty)|^2$ has a CIID structure, which enables the computation of several observables as in the other cases. 

Another interesting direction to explore is when the resetting is triggered not externally, 
as discussed so far, but by the system itself. For example, consider $N$ independent 
Brownian motions on the line starting at the origin at $t=0$. When one of the particles hits 
the level $L>0$ for the first time, all the $N$ particles are reset back to the origin 
simultaneously. For $N=1$, the system does not reach a stationary state \cite{BRR20} but it 
was shown recently that the system does reach a stationary state at 
long times for any $N>2$ 
with strong correlations between the particles, providing an example of DEC with 
event-driven stochastic drive~\cite{BMS26}. The stationary state for $N>2$ was also shown to have a CIID 
structure that enabled exact calculations of several observables despite the presence of 
strong correlations~\cite{BMS26}. A related issue is to study whether such event-driven 
simultaneous resetting expedites the search of a fixed target located at, say, 
$-x_0<0$~\cite{BMP25}. This was addressed recently in Ref. \cite{BMP25}  revealing
a rich and 
complex behavior for the mean first-passage time. Finally, very recently, the 
DEC and the associated CIID structure was found in models of catastrophes with small shocks 
under the diffusion approximation~\cite{Galla26}.

\noindent{\it Conclusion.} We have reviewed the recent progresses on DEC arising in classical and quantum noninteracting systems when they are subjected to a common fluctuating stochastic environment. These systems share two key properties: (i) strong all-to-all correlations emerge from the dynamics, and not from built-in interactions, (ii) the stationary state has a CIID structure that allows exact computations of physical observables. 
We also discussed recent experiments on trapped colloidal particles that have measured the DEC in the stationary state. This is a rapidly evolving area of research in out-of-equilibrium statistical physics with many open and challenging~problems. 

\noindent{\it Acknowledgments.} We warmly thank our collaborators on DEC: M. Barma, M. 
Biroli, A. Biswas, D. Boyer, S. Ciliberto, G. de Mauro, M. Kulkarni, H. Larralde, M. 
Mesquita, A. Nagar, A. Pal, A. Petrosyan, S. Sabhapandit. We acknowledge support from ANR Grant No. 
ANR-23-CE30-0020-01 EDIPS.

\end{document}